# Exploring Age-Related Patterns in Internet Access: Insights from a Secondary Analysis of New Zealand Survey Data

Edgar Pacheco
*Victoria University of Wellington – Te Herenga Waka,* edgar.pacheco@vuw.ac.nz

**Abstract:** About thirty years ago, when the Internet started to be commercialised, access to the medium became a topic of research and debate. Up-to-date evidence about key predictors, such as age, is crucial because of the Internet's ever-changing nature and the challenges associated with gaining access to it. This paper aims to give an overview of New Zealand's Internet access trends and how they relate to age. It is based on secondary analysis of data from a larger online panel survey with 1,001 adult respondents. The Chi-square test of independence and Cramer's V were used in the analysis. The study provides new evidence to understand the digital divide. Specifically, it uncovers a growing disparity in the quality of Internet connectivity. Even though fibre is the most common type of broadband connection at home, older adults are less likely to have it and more likely to use wireless broadband, which is a slower connection type. Additionally, a sizable majority of people in all age categories have favourable opinions on the Internet. Interestingly, this was more prevalent among older people, although they report an increased concern about the security of their personal information online. The implications of the results are discussed and some directions for future research are proposed.

**Introduction**
Since the commercialisation of the Internet in the 1990s (Grosse, 2020), access to the online medium has become a topic of research, policy intervention, and public discussion – with some voices arguing that it is a human right (Reglitz, 2020; Robinson, Schulz, Dunn, et al., 2020). Data from the World Bank estimates that the percentage of Internet users worldwide has progressively grown from 16 percent in 2005 to 66 percent in 2022 (World Bank, 2022). However, the overall increase in Internet access, particularly in Western countries (OECD, 2022; Poushter, 2016), does not necessarily imply that the access gap is closing. On the contrary, as some argue, it may be reproducing existing inequalities among, for instance, people from different socioeconomic backgrounds (DiMaggio & Garip, 2012; Zillien & Hargittai, 2009).

Of note, an early definition of Internet access centred on a binary differentiation regarding having or lacking physical access (van Deursen & van Dijk, 2015). While research on physical access is important, authors have called for a deeper understanding of the way people access the Internet. Nowadays, it is recognised that a comprehensive definition of Internet access includes, along with physical access, the understanding of Internet skills, motivations, attitudes, and expectations of getting physical access (Dijk, 2017) as well as the quality of the access (e.g. broadband access and affordability) (McClain et al., 2021; Prieger, 2013). Policy circles worldwide have started embracing this broader definition of Internet access as well (van Deursen & van Dijk, 2015). This has been the case in New Zealand.

A range of factors influencing Internet access (e.g., gender, ethnicity, income, educational status, employment, and living arrangements) has been reported in several studies (Büchi et al., 2016; ITP, 2022; König & Seifert, 2023; Pew Research Center, 2022). Attitudinal factors such as lack of perceived usefulness, trust, skills, and confidence, among other aspects, also have an impact on older adults' Internet access (Lips et al., 2020; McDonough, 2016). Moreover, evidence points out that differences also extend to those who are already Internet users (Zillien & Hargittai, 2009).

The role that age plays in Internet access has also been highlighted by prior research. For instance, a comparative study based on representative survey data from Sweden, the United States, Switzerland, the United Kingdom, and New Zealand found that, among all the sociodemographic factors analysed, age was the strongest predictor of Internet access and use (see Büchi et al., 2016). Studies also show that there are inequalities among different age groups. A comparative study found that, in each of the 40 countries investigated, younger adults aged between 18 and 34 reported higher rates of Internet use compared to their older counterparts (Poushter, 2016). In Germany, Huxhold (2020) found that older adults have a lower likelihood of Internet access compared to younger age groups. Age differences in Internet use also include skills. Older adults appear to be less experienced with touch screens on various devices such as phones, computers, and tablets (Volkom et al., 2014). Studies have also found that Internet access decreases as people become older (Diaz Andrade et al., 2021; Grimes & White, 2019). A recent article points out that age is associated with Internet users' preferences for and frequency of social media use, with younger adults aged 18-29 being more actively engaged with tools such as Facebook, Instagram, Snapchat, and TikTok compared those aged 50-64 and 65 and older (Pacheco, 2022).

While these age differences exist, longitudinal evidence also suggests that, in the last decade, older adults are the population segment with the highest growth in the use of the Internet and related technologies (Friemel, 2016). According to some studies, the COVID-19 pandemic has accelerated the trend (Lee, 2023; Sixsmith et al., 2022).

To summarise, as the number of people accessing the Internet grows, 'legacy' inequities such as those across age groups continue to exist, while new emerging challenges arise (Robinson, Schulz, Blank, et al., 2020). Thus, gathering and reporting updated evidence of trends regarding age, a key predictor of Internet access (Büchi et al., 2016), remains imperative to inform policy discussions and the development of targeted interventions aiming at increasing participation and inclusion as well as enhancing the opportunities and benefits of the Internet.

**The current study**

As described above, not having a connection is not the only challenge for Internet access. Both the complex and ever-changing nature of the Internet and its access mean that further nuance is required to understand access in the context of age. Uncertainty is particularly evident in New Zealand where there is limited quantitative data available (see Diaz Andrade et al., 2021; Grimes & White, 2019; InternetNZ, 2020). Addressing the evidence gap is even more critical considering that the New Zealand Government has reiteratively expressed its commitment to a digitally inclusive nation (DIA, 2019).

In order to explore trends in Internet access and its association with age, this paper uses the multifaceted model of Internet access suggested by van Deursen and van Dijk (2015) as a lens. Access to the online medium is understood in terms of the following facets: skills, usage, motivations, and material access. Specifically, the paper looks at frequent places of Internet access (home, work, and public Wi-Fi), as well as the type of broadband connection used at home. In addition, it looks at perceptions about issues related to Internet access (e.g., Internet speed, cost, and security). The findings provide a stocktake of trends and may inform discussion and policy interventions aiming at enhancing people's access to the online environment.

**Method/Design**

Data come from the New Zealand' Internet Insights, a modular online panel survey conducted by InternetNZ | Ipurangi Aotearoa. The survey explored different aspects related to people's interaction with digital technologies. Questions about Internet access comprised one of the modules and are the basis of this paper. However, while the survey's results provide general trends regarding access, inferential statistics were not applied to identified statistically significant group differences. For this reason, on request, the researcher asked InternetNZ | Ipurangi Aotearoa for the anonymised raw survey data and data dictionary to perform secondary data analysis and explore in more detail the role of age regarding specific aspects of Internet access.

Secondary data analysis is the analysis of any data in order to answer a research question(s) different to the questions(s) for which the data was originally gathered (Vartanian, 2010). The approach not only provides researchers with an alternative when time and resources are limited but also an opportunity when research activities are disrupted due to unexpected events such as the COVID-19 pandemic (Spurlock & Spurlock, 2020). Secondary data analysis requires applying the same basic research principles as any research method using primary data (Johnston, 2017). Reusing data for additional analysis is a well-established approach in the social sciences and health sciences (Cole & Trinh, 2017; Smith, 2008).

The survey was administered online to 1,001 New Zealanders who are Internet users aged 18 and older by Kantar Public's online consumer panels. The market research company used a combination of pre-survey quotas and post-survey weighting to ensure results are representative of all New Zealanders by key demographics such as age. Online surveys are not only cost-effective and easier to administer but also their use as a technique for data collection is growing in the social sciences and policy research (Lehdonvirta et al., 2021; Sue & Ritter, 2012). In New Zealand, online surveys have been shown to be useful for exploratory research on aspects related to the opportunities and risks of digital technologies (Pacheco, 2022; Pacheco & Melhuish, 2018, 2020).

Consent to take part in the survey was obtained from all participants, who had the right to withdraw from it at any time.

Fieldwork was conducted from 3 to 17 November 2021. At the time of data collection parts of the country (Auckland, Upper Northland, and Waikato) were in Level 3 of New Zealand's COVID-19 Alert System while the rest were in Level 2. Details of what measures and responses the system comprised to deal with the pandemic can be found on New Zealand Government websites (see New Zealand Government, 2022).

The maximum margin of error on the total group (n=1,001) is +3.1 percent at the 95 percent confidence interval.

**Sample**
The distribution of the sample was as follows. In terms of gender, females represented 51.8 percent while males represented 48.1 percent. Only one participant identified as gender diverse (0.1 percent). Unfortunately, due to the small sample size, they were excluded from the analysis. The age distribution was: 18-19-year-old participants (0.9 percent), 20-24-year-olds (9.3 percent), 25-29-year-olds (11.9 percent), 30-34-year-olds (7.7 percent), 35-39-year-olds (8.7 percent), 40-44-year-olds (8.9 percent), 45-49-year-olds (8.0 percent), 50-54-year-olds (8.1 percent), 55-59-year-olds (9.1 percent), 60-64-year-olds (7.6 percent), and 65-year-olds and over (19.9 percent).

**Measures**
*Age group*
Age group was the independent variable. For analysis, participants were aggregated in three age group categories. Younger adults were encompassed by those who indicated their age was between 18 and 39 years old. Middle-aged adults were represented by those aged between 40 and 59. The third age group, older adults, included those aged 60 years old and over.

*Internet access*
Three of the questions in the survey looked at the frequency of Internet access in specific places, i.e., at home, at work, and via public Wi-Fi. These questions were related to usage and material access in van Deursen and van Dijk's (2015) multifaceted Internet access approach. For each question the following options were included: 'Once a day or more often', 'Two or three times a week', 'Once a week', 'Two or three times a month', 'Once a month', 'Less than once a month', and 'Never'.

The survey also included another question related to material Internet access (van Deursen & van Dijk, 2015). In this question, participants were asked about the type of Internet connection they have at home. The responses for this question included the following: 'Fibre (any connection)', 'ADSL', 'VDSL', 'Wireless broadband', 'Satellite', 'Don't know', and 'None'.

Four questions in the survey were related to motivational access (van Deursen & van Dijk, 2015). Participants were asked about their level of concern about issues related to Internet access in New Zealand. To this end, three Likert-type items were included based on the following statements: 'Access is poor in remote areas', 'Cost of Internet access is high', and 'slow speed of Internet in New Zealand'. For each of these statements, participants were asked to think about society as a whole, rather than anything they may or may not have personally experienced. They could choose from the following scale: 'Extremely concerned', 'Very concerned', 'A little bit concerned', 'Not very concerned', 'Not at all concerned', and 'I don't know'.

Participants were also asked about their attitudes towards the Internet: 'There are positives and negatives to the Internet, but overall do you think the positives outweigh the negatives?'. Respondents were able to choose from 'Yes', 'No', and 'I don't know'. The survey also included two questions regarding safety online. One question asked: 'How concerned are you about the security of your own personal details when you use them on the Internet?' The response scale included the following items: 'Extremely concerned', 'Very concerned', 'A little bit concerned', 'Not very concerned', 'Not at all concerned', and 'Don't know'. The second question asked participants whether their level of concern had changed over the previous 12 months. The options to answer the question were 'Yes', 'No', and 'I don't know'.

The survey did not include questions related to self-assessment of participants' own Internet skills which are usually used to measure skills (van Deursen & van Dijk, 2015).

**Data analysis**
Categorical variables are presented in contingency tables as frequencies and percentages. To test significance of association, we used the Chi-square test of independence, a statistical procedure applied to explore whether two variables are likely to be associated (McHugh, 2013). The procedure also offers thorough details on which categories explain any disparities discovered (McHugh, 2013). Key assumptions of the Chi-square test require having two categorical variables, independent observations, mutually exclusive categories, and expected frequencies greater than 5 in at least 80 percent of cells (McHugh, 2013). The data used for the current work meet these assumptions. While it is a frequently-used procedure in the social sciences (Turhan, 2020), the Chi-square test of independence cannot establish a causal relationship between two variables, only whether an association exists. Another limitation is that it does not provide a measure of the magnitude of the association. Considering the latter, we conducted Cramer's V, which is the most commonly used effect size measurement for the Chi-square test of independence and is applied when the contingency table is larger than 2x2. Cramer's V measures how strongly two categorical variables are associated (Kearney, 2017). However, a limitation of Cramer's V is that it tends to produce relatively low correlation measures (McHugh, 2013). We use Cohen's (1988) recommended interpretation to observe the strength of association. Statistical significance was determined at $p < .05$. We also collapsed categories into larger categories to analyse the observed counts regarding places of Internet access (home, work, and public Wi-Fi), and levels of concern about issues related to Internet access. Collapsing data across observations is a common practice in survey research (Manor et al., 2000). All analyses were performed using the Jamovi software, version 2.3 (The jamovi project, 2023).

**Results**

The Chi-square test of independence was performed to examine the association between the frequency of Internet access at home and age group. The association between these variables was significant, χ² (6, N = 1001) = 51.7, p < .001. However, the effect size for this finding, Cramer's V, was small, .16. As can be seen in Table 1, while rates of frequent Internet access at home were significantly high among all age groups, middle-aged adults were more likely to connect at home daily than younger and older adults.

On the other hand, as can be seen in the frequencies cross-tabulated in Table 1, there was a significant association between the frequency of Internet access at work and age group, χ² (8, N = 1001) = 232, p < .001. The effect size, Cramer's V, was large, .34. Middle-aged adults followed by younger adults were more likely to frequently access the Internet at work (once a day or more often) compared to older adults.

We also looked at the association between Internet access via public Wi-Fi and age group. The Chi-square test of independence revealed a significant association between these two nominal variables, χ² (8, N = 1001) = 118, p < .001. The effect size for this result, Cramer's V, was medium, .24. Frequent access to the Internet via public Wi-Fi (once a day or more often) was higher among younger adults than middle-aged and older adults. Furthermore, older adults were significantly more likely to never use public Wi-Fi to connect to the Internet (see Table 1 for details).

**Table 1**

*Frequency of Internet Access at Home, Work, and Public Wi-Fi by Age Group*

| Place of access | Age group | | | | | | | |
|---|---|---|---|---|---|---|---|---|
| | Younger adults | | Middle-aged adults | | Older adults | | Total | |
| | n | % | n | % | n | % | n | % |
| **Access at home *** | | | | | | | | |
| Daily | 343 | 89.1 | 334 | 97.9 | 251 | 91.3 | 928 | 92.7 |
| Weekly or more often | 37 | 9.6 | 7 | 2.1 | 13 | 4.7 | 57 | 5.7 |
| Monthly or more often | 4 | 1.0 | 0 | 0.0 | 0 | 0.0 | 4 | 0.4 |
| Never | 1 | 0.3 | 0 | 0.0 | 11 | 4.0 | 12 | 1.2 |
| **Access at work *** | | | | | | | | |
| Daily | 284 | 73.8 | 273 | 80.1 | 103 | 37.5 | 660 | 65.9 |
| Weekly or more often | 56 | 14.5 | 28 | 8.2 | 19 | 6.9 | 103 | 10.3 |
| Monthly or more often | 7 | 1.8 | 2 | 0.6 | 4 | 1.5 | 13 | 1.3 |
| Less than once a month | 2 | 0.5 | 3 | 0.9 | 6 | 2.2 | 11 | 1.1 |
| Never | 36 | 9.4 | 35 | 10.3 | 143 | 52.0 | 214 | 21.4 |
| **Access via public Wi-Fi *** | | | | | | | | |
| Daily | 52 | 13.5 | 23 | 6.7 | 9 | 3.3 | 84 | 8.4 |
| Weekly or more often | 119 | 30.9 | 79 | 23.2 | 24 | 8.7 | 222 | 22.2 |
| Monthly or more often | 86 | 22.3 | 78 | 22.9 | 49 | 17.8 | 213 | 21.3 |
| Less than once a month | 79 | 20.5 | 97 | 28.4 | 89 | 32.4 | 265 | 26.5 |
| Never | 49 | 12.7 | 64 | 18.8 | 104 | 37.8 | 217 | 21.7 |

Note: * $p < .001$. 'Don't know' not included in analysis.

The association between the type of broadband access at home and age group was also analysed via the Chi-square test of independence. These two nominal variables showed a significant but small association, $\chi^2$ (10, N = 960) = 31.4, p < .001, *V* = .12. Fibre was the most common type of Internet connection among all age groups, however, older adults were less likely to have this at home and much more likely to use wireless broadband than younger adults and middle-aged adults (see Table 2).

**Table 2**

*Type of Internet Connection at Home by Age Group*

| Type of Internet connection at home * | Age group | | | | | | | |
|---|---|---|---|---|---|---|---|---|
| | Younger adults | | Middle-aged adults | | Older adults | | Total | |
| | n | % | n | % | n | % | n | % |
| Fibre (any connection) | 249 | 67.5 | 217 | 66.4 | 152 | 57.6 | 618 | 64.4 |
| ADSL | 36 | 9.8 | 20 | 6.1 | 15 | 5.7 | 71 | 7.4 |
| VDSL | 14 | 3.8 | 20 | 6.1 | 11 | 4.2 | 45 | 4.7 |
| Wireless broadband | 60 | 16.3 | 64 | 19.6 | 83 | 31.4 | 207 | 21.6 |
| Satellite | 4 | 1.1 | 4 | 1.2 | 0 | 0.0 | 8 | 0.8 |
| Other | 6 | 1.6 | 2 | 0.6 | 3 | 1.1 | 11 | 110 |
| Total | 369 | 12.7 | 327 | 18.8 | 264 | 37.8 | 960 | 21.7 |

Note: * p < .001. 'Don't know' (n=41) not included in analysis.

As indicated in the methodology section, the survey also included three Likert-type items looking at the level of concern among the participants about issues related to Internet access in New Zealand. For attitudes regarding poor Internet access in remote areas (see Table 3), older adults were more likely to report being very or extremely concerned compared to the other two age groups, χ² (4, N = 974) = 15.3, p = .004 although Cramer's V = .08 showed a weak effect size.

Regarding the level of concern about the cost of Internet, the Chi-square test of independence showed that there was no significant association between this nominal variable and age group, χ² (4, N = 985) = 3.83, p = .43. Similarly, no significant difference was found for the level of concern about slow Internet speed by age group, χ² (4, N = 984) = 7.58, p = .10. See Table 3.

**Table 3**

*Perceptions about Current Issues Related to Internet Access by Age Group*

|  | Age group | | | | | | | |
| --- | --- | --- | --- | --- | --- | --- | --- | --- |
| **Statement about Internet access** | **Younger adults** | | **Middle-aged adults** | | **Older adults** | | **Total** | |
|  | n | % | n | % | n | % | n | % |
| **Poor access in remote areas *** | | | | | | | | |
| Very/extremely concerned | 169 | 44.9 | 164 | 49.2 | 146 | 55.1 | 479 | 49.2 |
| A little bit concerned | 132 | 35.1 | 98 | 29.4 | 90 | 34.0 | 320 | 32.9 |
| Not very/not at all concerned | 75 | 19.9 | 71 | 21.3 | 29 | 10.9 | 175 | 18.0 |
| **Cost of Internet access is high** | | | | | | | | |
| Very/extremely concerned | 137 | 36.1 | 131 | 38.8 | 98 | 36.6 | 366 | 37.2 |
| A little bit concerned | 143 | 37.7 | 132 | 39.1 | 115 | 42.9 | 390 | 39.6 |
| Not very/not at all concerned | 99 | 26.1 | 75 | 22.2 | 55 | 20.5 | 229 | 23.2 |
| **Slow Internet speed in New Zealand** | | | | | | | | |
| Very/extremely concerned | 145 | 38.4 | 151 | 44.7 | 94 | 35.1 | 390 | 39.6 |
| A little bit concerned | 119 | 31.5 | 104 | 30.8 | 98 | 36.6 | 321 | 32.6 |
| Not very/not at all concerned | 114 | 30.2 | 83 | 24.6 | 76 | 28.4 | 273 | 27.7 |

Note: * $p < .004$. 'Don't know' not included in analysis.

As mentioned, participants were asked to think about whether, overall, the positives of the Internet outweigh the negatives. In this respect, age was found to be significantly associated with this variable, $\chi^2 (2, N = 906) = 9.41$, $p = .009$, although Cramer's $V = .10$ showed a weak effect size. A large majority in each age group answered the question positively, with older adults slightly more likely to say 'Yes' than the other younger adults and middle-aged adults (see Table 4).

**Table 4**

*Attitudes about the Internet by Age Group*

| Positives overweight the negatives * | Age group | | | | | | Total | |
|---|---|---|---|---|---|---|---|---|
| | Younger adults | | Middle-aged adults | | Older adults | | | |
| | n | % | n | % | n | % | n | % |
| Yes | 310 | 91.4 | 296 | 94.9 | 248 | 97.3 | 854 | 94.3 |
| No | 29 | 8.6 | 16 | 5.1 | 7 | 2.7 | 52 | 5.7 |
| Total | 339 | 100.0 | 312 | 100.0 | 255 | 100.0 | 906 | 100.0 |

Note: * p = .009. 'Don't know' (n=95) not included in analysis.

The association between the level of security concern about participants' own personal details when they use them on the Internet and age was also tested. The Chi-square test of independence showed a significant association, $\chi^2$ (6, N = 1,001) = 23.2, p < .001, although Cramer's V = .10 showed a weak effect size. Older adults, followed by middle-aged adults, were more likely to say they are very or extremely concerned. Regarding an increase in the level of concern over the security of personal details online in the last year and the link with age, a significant but weak association was found, $\chi^2$ (4, N = 990) = 10.0, p = .040, V = .07. A large majority across all three age groups indicated that their level of concern did not change in the last 12 months. This was higher among younger adults. Also, older adults were more likely to say that their concern has increased (see Table 5).

**Table 5**

*Attitudes towards Security of Personal Details Online by Age Group*

|  | Age group |  |  |  |  |  |  |  |
|---|---|---|---|---|---|---|---|---|
|  | Younger adults | | Middle-aged adults | | Older adults | | Total | |
|  | n | % | n | % | n | % | n | % |
| **Level of concern about security of personal details online *** | | | | | | | | |
| Very/extremely concerned | 187 | 48.6 | 185 | 54.3 | 152 | 55.3 | 524 | 52.3 |
| A little bit concerned | 160 | 41.6 | 132 | 38.7 | 99 | 36.0 | 391 | 39.1 |
| Not very/not at all concerned | 36 | 9.4 | 23 | 6.7 | 14 | 5.1 | 73 | 7.3 |
| I don't use personal details online | 2 | 0.5 | 1 | 0.3 | 10 | 3.6 | 13 | 1.3 |
| **Change in level of concern about security of personal details online ** ** | | | | | | | | |
| Increased | 125 | 32.2 | 127 | 37.4 | 112 | 40.9 | 364 | 36.8 |
| Decreased | 7 | 1.9 | 2 | 0.6 | 0 | 0.0 | 9 | 0.9 |
| Stayed the same | 244 | 64.9 | 211 | 62.1 | 162 | 59.1 | 617 | 62.3 |

Note: * $p < .001$. ** $p = .040$.

## Discussion and conclusion

The purpose of this paper was to explore trends in Internet access in New Zealand and their association with age based on survey data. The findings reveal that, despite nearly three decades since the commercial inception of the Internet, there are still age-related differences among those who are already Internet users, particularly in terms of the quality of broadband access. The findings are discussed below.

Regarding the types of broadband connectivity, and similar to Diaz Andrade and colleagues (2021), we found that fibre is the most common type of Internet connection in New Zealand homes. Our results show that nearly two-thirds of participants in our sample connect online via fibre – this was about 5 percentage points lower than the result from the aforementioned study (see Diaz Andrade et al., 2021).

Furthermore, an important contribution of the current study is that it found statistically significant differences in the types of connection at home, with younger adults followed by middle-aged adults being more likely to adopt fibre. However, older adults are less likely to have this type of faster Internet connection, about 10 percentage points lower than younger adults. It is also interesting that 31.4 percent of older adults have wireless broadband at home – a type of connection that delivers lower download/upload speeds and more frequent dropouts than fibre, has the highest latency of all technologies apart from satellite, and offers, on average, slower speeds than VDSL (Commerce Commission, 2022). The rate reported by older adults nearly doubles younger adults' use of wireless broadband at home.

The results described above show an age gap in the quality of broadband access among New Zealanders. It is not only that older adults are less likely to use the Internet (Grimes & White, 2019), navigate the online environment less often, and perform fewer activities than their younger counterparts (Mariano et al., 2021; Pacheco, 2022); they are also, as the present study shows, less likely to have a faster broadband connection at home. Access to high-speed broadband has become more important with the COVID-19 pandemic, particularly for older adults, as people relied on the online medium to work from home, access health care services, connect with family and friends, apply for public assistance, and perform basic activities such as buying groceries (McClain et al., 2021). Access to high-speed broadband among older adults is important as it is associated with greater digital skills (Hargittai & Dobransky, 2017) and, in the context of the pandemic, perceived quality of life (Mohan & Lyons, 2022; Wallinheimo & Evans, 2021). While research and commentary has pointed out the limited access to broadband connection among older adults (Pew Research Center, 2021), the current study also uncovers age disparities in the quality of connection among those who are already broadband users. The difference discussed here contributes to our understanding of age-related challenges for digital equity and sheds light on a component of the broadband divide (Riddlesden & Singleton, 2014) that, so far, has centred on inequalities between rural and urban areas.

Thus, continued efforts are needed to address the age gap in the quality of broadband access. In New Zealand, current initiatives towards Internet access include community-based digital training (i.e., Better Digital Future for Seniors), subsidised prepaid broadband access offering 35GB of data (i.e., Skinny Jump), and access to technical support to help older adults use apps on smartphones, tablets, and computers (i.e., Appy Seniors). While these initiatives play a role to reduce the gap, they can be complemented by other responses, such as expanding affordable access to high-speed broadband connection as, for some older adults, the cost of high-speed Internet makes it unaffordable. In addressing the age gap in the quality of broadband access, older adults will have better opportunities for equal access to government information, health services, banking, shopping, and entertainment, especially as we transition to the post-COVID-19 pandemic era.

Furthermore, the age difference in broadband access quality reported here needs to be investigated further in regard to its relationship with other disparities affecting digital inclusion in New Zealand. In terms of Internet access, there is some international evidence that social, economic, and geographic inequalities intersect with age. For instance, older adults with a greater degree of education and wealth have better Internet skills (Hargittai & Dobransky, 2017; Kottorp et al., 2016). However, older adults in long-term care facilities who have physical or cognitive disabilities are barred from using digital technologies, including the Internet (Seifert et al., 2021). When compared to those living in urban areas, older adults in rural regions have lower levels of Internet access and activities (Lee et al., 2020). Equitable participation in society through digital technologies has been a major policy concern in the country (see digital.govt.nz, 2022). However, in order to successfully address age-related disparities in Internet access, future research and policy initiatives must consider how other current inequalities overlap.

Regarding perceptions of the Internet, a large majority of participants, over 9 out of 10, consider that the positives of the online medium outweigh the negatives. Older adults in our sample tend to have a more favourable view of the Internet compared to younger age groups. One possible explanation for this result may be the COVID-19 pandemic, as fieldwork was conducted when the country was facing restrictions and adapting to the life changes the pandemic imposed. This claim appears to be supported by recent research, which reveals that during COVID-19, a large majority of people–including older adults– perceived the Internet as crucial or significant (McClain et al., 2021). Additionally, studies show that positive attitudes towards the Internet are associated with increased rates of Internet usage, online shopping, and psychological well-being among older adults (Iyer & Eastman, 2006; Zambianchi & Carelli, 2018). The results of the current study emphasise the significance of the Internet based on age characteristics, but in order to fully comprehend older individuals' experiences with the online medium, we need more specific data regarding their behaviours and attitudes towards it.

When looking at the results about the security of personal details online, just over half of participants indicated they were extremely or very concerned about the security of their personal details when used on the Internet. Moreover, a third of participants said their level of concern on this matter had increased in the past year. Older adults were slightly more likely to be concerned, and to report an increase in their level of concern in this regard. While there are discernible age disparities, this is not an easy relationship to understand. Branley-Bell et al. (2022), for example, found that, compared to younger people, older adults are more likely to create secure passwords, show proactive risk awareness, and regularly install updates, although they are less likely to secure their devices. A New Zealand-based study shows that most older adults (aged 50 and over) feel reasonably confident in their ability to keep their digital information confidential and secure (New Zealand Seniors, 2022). However, in contrast to this behaviour and perceptions, older adults are increasingly vulnerable to financial scams, phishing attacks, and consumer fraud online (Burnes et al., 2017; Grilli et al., 2021) and other online risks (Melhuish & Pacheco, 2021). While our findings supplement growing evidence regarding aging and cybersecurity (Morrison et al., 2021), a more comprehensive survey will provide deeper understanding on the topic.

Regarding frequent places of Internet access, comparison of the findings with those of other studies (Grimes & White, 2019; McClain et al., 2021) confirms that home is where a large majority of people, about 9 in 10, connect daily to use the Internet. When looking at age, there were differences, particularly between middle-aged and younger adults – about 8 percentage points. Meanwhile, with respect to frequent Internet access at work, two-thirds indicated they use the Internet daily in their workplace. Rates were higher among middle-aged adults followed by younger adults, but significantly lower for older adults. This finding was expected as middle-aged and younger adults mainly comprise the country's workforce (Stats NZ, 2021). The findings regarding Internet access via public Wi-Fi also deserve attention. In this regard, connecting online via this means is, overall, much less frequent among all age groups compared to other places of access. Despite this, younger adults were more likely to use public Wi-Fi daily or weekly than middle-aged and older adults. Meanwhile, older adults were particularly less likely to connect online through public Wi-Fi, not only on a daily but also a weekly basis. Research has warned of the privacy and security risks of public Wi-Fi networks (Cheng et al., 2013; Lotfy et al., 2021), and the evidence points out that the desire to conserve mobile data allowance, gender, and educational level are related to people's use of these facilities (Sombatruang et al., 2018). While more research is needed to understand whether age is also a predictor, the current work is a first step in this direction as it shows that there is an association with younger people tending to use publicly available Wi-Fi networks more often to connect online.

When looking at the three Likert-type questions, the results are mixed. The data shows that the level of concern about poor Internet access in remote areas was associated with age. In this respect, nearly half of participants say they are extremely or very concerned about this, with a higher rate among older adults (55.1 percent). However, while 37.2 percent of participants expressed their concern about the cost of Internet access, and 39.6 percent did so in terms of slow Internet speed, when compared with age, the results are inconclusive. In this respect, the Chi-square test found no statistically significant differences between these two Internet-related issues and age groups.

Finally, this paper is a stocktake of trends in Internet access in New Zealand. A natural progression of it would be to further explore the extent and nature of Internet access using a smartphone as, according to research, it is one of the main reasons for not having a high-speed Internet connection at home (Anderson, 2019). Another potential avenue for research is to investigate in more detail differences in Internet access and use among older adults as they represent a heterogeneous group and evidence suggests that their online activities also differ (Robinson, Schulz, Blank, et al., 2020).

**Limitations**
While this paper was able to shed some additional light on Internet access in the New Zealand context and its association with age, it is not without limitations. As the survey was completed online, it was not possible to gather data from those who, for different reasons, are not Internet users. Despite this, we are confident in the findings as the focus of this paper was on those who have access to the Internet. In addition, the sampling strategy and weighting allowed for a representative sample of those already-connected adult New Zealanders in terms of age.

Another limitation is that the findings in this paper are restricted by the original survey design. They represent a stocktake of trends. Thus, a larger follow-up study will be needed.

In addition, the Chi-square test of independence is a robust statistical test to analyse the association between categorical variables (McHugh, 2013); however, we acknowledge that it does not permit determination of causality. Despite this, the exploratory character and the type of variables used for the paper make the Chi-square test the best analytical choice.

Finally, another limitation is that the data were collected at one point in time; thus, the findings represent a snapshot of participants' experiences. To understand how trends of Internet access change over time, longitudinal evidence is needed.